\documentstyle[12pt,aps,prc,epsf]{revtex}
\begin{document}
\begin{center}
{\Large \bf Excitons and Polarons in Quantum Wells}\\
\bigskip
{\large B.\ Gerlach$^{a)}$, M.\ A.\ Smondyrev$^{b)}$} \\
\date{}
\smallskip
{\it $^{a)}$Institut f\"ur Physik, Universit\"at Dortmund,
44221 Dortmund, Germany \\
$^{b)}$Bogoliubov Laboratory of Theoretical Physics,
JINR, 141980 Dubna, Russia}
\end{center}

\section{General Discussion}

The purpose of this article is to analyze the dependence of the energy
of an elementary excitation on the strength of the confinement
potential, which exists in a planar semiconductor heterostructure. Due
to the fascinating technological progress in the field of man-made
structures, it has become possible to fabricate e.g. quantum wells of a
widely varying shape. It is an interesting theoretical task to discuss
the excitation spectrum of such semiconductor structures as function of
the tunable parameters, such as well width, well height, etc..
Concerning the excitations of interest, we concentrate on
particle-phonon systems, the particles being electrons or holes. The
simplest example is that of a single polaron, that is an electron,
coupled to a certain branch of lattice vibrations.  Another example is
that of a polaronic exciton, that is an electron-hole pair, coupled to
phonons. Whereas the latter one is important to characterize optical
properties, the former one has direct implications for the transport
behavior of the materials of interest.

We assume that the interface(s)-induced confinement can be mimicked by a
simple potential $V_n(z_n)$, $n$ being the particle number, $z_n$ the
corresponding coordinate (the growth direction of the heterostructure will
always be assumed as $z$-direction). Explicit forms of $V_n(z_n)$ may
be rectangular wells, parabolas etc.. In addition, we suppose translation
invariance to hold within the $xy$-plane. We remark that effects as
surface roughness would destroy this property and could lead to the
appearanc of new phenomena (e.g. localized states).

In the following equation, we define the class of models under discussion:

\begin{eqnarray} \label{model}
H:&=&\frac{1}{2}\sum_{n=1}^{N} {\bf{p}}_n {m_n}^{-1} {\bf{p}}_n +
   U({\bf{r}}_1,..,{\bf{r}}_N) + \nonumber \\
&&   \sum_{\bf{k}}\hbar\, {\omega}_{\bf{k}}\,a_{\bf{k}}^{+}a_{\bf{k}}+
   \frac{1}{\sqrt{V}}\sum_{n=1}^{N}\sum_{\bf{k}}\left(g_{\bf{k},n}
   \,e^{\,i\,\bf{k}\bf{r}_{n}}\,a_{\bf{k}}+h.c.\right) \nonumber \\
   [.4 cm]
&=&: H_{el} + H_{ph} + H_{int} .
\end{eqnarray}


\noindent
The nomenclature is self explaining. The quantity $U(\bf{r}_1,\bf{r}_2)$
is to contain the confinement potentials as well as the particle
interaction:

\begin{equation} \label{potential}
U({\bf{r}}_1,..,{\bf{r}}_N):= \sum_{n=1}^{N}\,V_n(z_n) +
\frac{1}{2}\sum_{
\begin{array}{l}
\scriptsize n,n'=1, \\[-2mm]
\scriptsize n\neq n'
\end{array}
}^{N}\,V_{n,n'}({\bf{r}}_n,{\bf{r}}_{n'}),
\end{equation}
\noindent
where $V_{n,n'}$ has to be calculated as potential
energy of particle $n$, exposed to the electrostatic potential of
particle $n'$. Because of the boundary conditions,
$V_{n,n'}({\bf{r}}_n,{\bf{r}}_{n'})$ itself is not translation invariant
(see e.g. Ref. \cite{Takagahara}). The particle-phonon coupling is of
Fr\"ohlich type. The most prominent example to be used here is
that of a coupling to (LO)-phonons.

The model has two relevant limiting cases, which should be reproduced by
any theory. Let the maximum of the well widths be $L$ and the minimum
$L'$. If $L'$ tends to infinity, the confinement is irrelevant and the
energy spectrum of $H$ is that of a three-dimensional well-material
excitation. If $L$ tends to zero, the (finite height) well is irrelevant,
leaving us with the spectrum of a three-dimensional barrier-material
excitation. The behaviour for intermediate values of the well widths
can qualitatively be discussed as follows. Varying $L,L'$ from sufficiently
large values to smaller ones, the binding energy should increase due to
the higher Coulomb correlation (for instance, the reader should recall
that the energy of the two-dimensional hydrogen ground state is four times
larger than that of a three-dimensional one). When $L,L'$ become smaller
and smaller, the ground-state wave function will more and more effectively
tunnel into the barrier material --- the energy approaches the barrier limit.

Thus, we might expect a maximum of the binding energy to appear at intermediate
values of $L,L'$. It was a controversely discussed question whether or
not this maximum appears at relevant (that is not too small) values of
$L$.  The answer to this question might be not the same for different
systems.

\section{Polarons}

The physics of polarons, confined to quantum wells, passed a few stages, and
it is not possible to present here even a brief list of references. In
particular, it was found that different phonon modes contribute to the
polaron binding energy --- confined bulk 2phonons inside the well,
interface phonon mode and half-space bulk phonon mode in the barrier.
We cite only papers\cite{Hai,Shi} concerning polarons confined to a
finite rectangular potential (one layer heterostructure) where contribution
of all phonon modes were taken into account. Anyway, there are problems to
be addressed while dealing with multilayered heterostructures. Namely, we
have to answer the following questions:

1) How to deal with multilayered heterostructures? The total number of
phonon modes becomes too large to make numerical calculations even with
modern computers. Besides, a multilayered heterostructure can generate
a confining potential of rather complicated form, not only the rectangular one.

2) How to deal with mass- and dielectric mismatches in different layers?
The polaron effective mass $m(z)$, the electron-phonon coupling constant
$\alpha(z)$ and the phonon dispersion law do depend on a layer, that is, on
the electron position. To glue solutions in different layers seems
to be a cumbersome job.

To tackle these problems we suggest specific approximations, which will
briefly be indicated here.

\begin{itemize}
\item A multilayered $GaAs/Al_xGa_{1-x}As$ heterostructure is considered as
an {\em effective medium}. Its mean parameters are to be defined by
averaging over different layers according to the way they enter the
Hamiltonian. \\[-7mm]
\item The bulk phonon mode only inhabits an effective medium with mean
characteristics.
\end{itemize}

We specify the electronic part of the Hamiltonian:

\begin{eqnarray}
H_{el} = H_{el,\parallel} + H_{el,\perp} =
{\vec p_{\parallel}^{\,2}\over 2m} + {p_z^{\,2}\over 2m} + V(z) ,
\label{eq2.02}\end{eqnarray}

\noindent
The \underline{mean electron band mass} $m$ is defined by the equation
\begin{eqnarray}
H_{el,\perp}\psi_1 = E_1 \psi_1, \qquad
{1\over m} = \int\limits_{-\infty}^{\infty} dz\,{|\psi_1(z)|^2\over m(z)} ,
\label{eq2.03}\end{eqnarray}

\noindent
where $\psi_1(z), E_1$ are the ground state wave function and the energy
for the electron motion in $z$ direction. As $\psi_1$ and $E_1$ depend
on $m$, we actually have the system of two equations (\ref{eq2.03})
to calculate the mean band mass $m$.

The free LO-phonon Hamiltonian reads as follows:

\begin{eqnarray}
H_{ph} = \hbar \omega_{\mbox{\tiny LO}}
\sum\limits_{\vec k} a^{\dag}_{\vec k}a_{\vec k} \ \ , \qquad
\omega_{\mbox{\tiny LO}}= \int\limits_{-\infty}^{\infty} dz\,
\omega(z)\,|\psi_1(z)|^2 .
\label{eq2.05}\end{eqnarray}

\noindent
As $m$ is found already, we define here the \underline{mean phonon frequency}
$\omega_{\mbox{\tiny LO}}$. Note that in this paper we are not interested
in processes of emission, absorption or scattering of phonons. Instead we
concentrate on virtual phonons in a cloud around an electron. Subsequently,
the properties of the {\em effective} phonons do depend
on the position of the electron as it follows from Eq.~(\ref{eq2.05}).

In the same way we define the effective electron-phonon
interaction Hamiltonian in the standard Fr\"ohlich form with
the \underline {mean Fr\"{o}hlich coupling constant} $\alpha$:

\begin{eqnarray}
\sqrt{\alpha}
= \int\limits_{-\infty}^{\infty} dz\, |\psi_1(z)|^2\,{\omega(z)\over
\omega_{\mbox{\tiny LO}}} \left( \alpha(z)
\sqrt{m\omega_{\mbox{\tiny LO}}\over m(z)\omega(z) } \right)^{1/2}  .
\label{eq2.07}\end{eqnarray}

Evidently, this model belongs to the class defined in Eq.~(\ref{model})
As examples we studied 1) a one-layer heterostructure described
by a rectangular confining potential

\begin{eqnarray}
V(z) = \left\{
\begin{array}{ll}
0,\, & |z| \leq L/2  \\
V_0, & |z| > L/2
\end{array} \right.  .
\label{eq3.01}\end{eqnarray}

\noindent
(the $z$-dependence of the masses and dielectric parameters is completely
analogous) and 2) a multilayered heterostructure corresponding to the
Rosen-Morse potential

\begin{eqnarray}
V(z) &=& V_0\,{\rm tanh}^2\left({z\over L_{RM}}\right)  .
\label{eq4.01}\end{eqnarray}

We use perturbation theory in powers of $\alpha$ for both potentials,
but in the first case we perform the summation over all virtual states
while in the case of the Rosen-Morse potential the Green function
(see \cite{Kleinert,Leschke}) can be used. To compare results for the
Rosen-Morse and the rectangular potentials, an effective width $L$ of
the Rosen-Morse potential has to be found. We define it as the width
of a rectangular potential of the same height $V_0$ with the same
ground-state energy. The dependence $L(L_{RM})$ can then be calculated.
The parametrization for experimental data concerning $GaAs/Al_x Ga_{1-x} As$
heterostructure is based on the results reported in Ref.~\cite{Adachi} with
some modifications, which are discussed in our paper\cite{rosen}.
Actually we use the dependence of the parameters on
the $Al$ mole fraction $x$ which depends in turn on the coordinate
$z$ via the relation
$V(z) = 600 \cdot (1.155 x + 0.37 x^2 )\ {\rm meV}$.
The confining potential $V(z)$ being given, we know the dependence
$x(z)$ and, subsequently, the values of the parameters $\alpha, m, \omega$
at each point of the heterostructure which are averaged then following
Eqs.~(\ref{eq2.03}), (\ref{eq2.05}) and (\ref{eq2.07}).

  \begin{figure}[h]
  \hspace*{-0.5cm}
  \epsfysize=10in \epsfbox{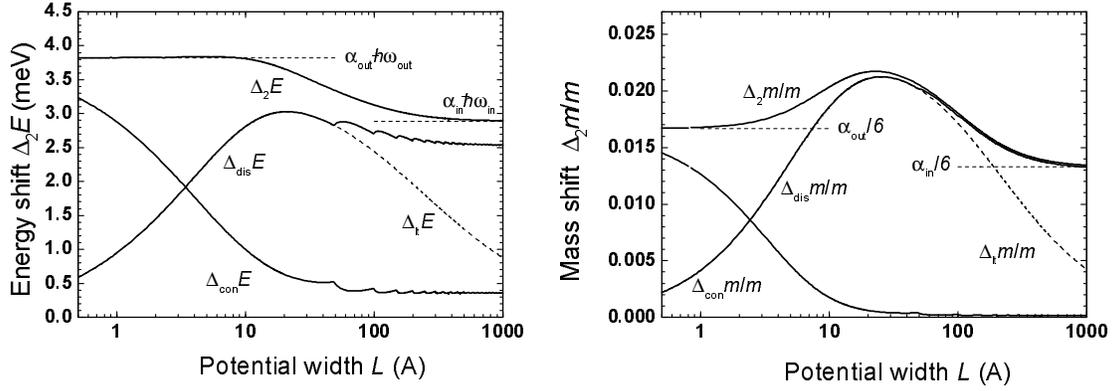}
  \vspace*{-19.0cm}
  \caption{The polaron binding energy and the effective mass
in the rectangular potential. Contributions of the discrete
$\Delta_{dis}E \ (\Delta_{dis}m/m)$ and
continuous $\Delta_{con}E \ (\Delta_{con}m/m)$
spectrum are shown as well as the so called leading
term approximation $\Delta_{lt}E \ (\Delta_{lt}m/m)$
when only the ground state is taken into account
as an intermediate virtual state (dashed line).}
  \label{bog-fig1}\end{figure}

\begin{figure}[b]
  \hspace*{-1.0cm}
  \epsfysize=10in \epsfbox{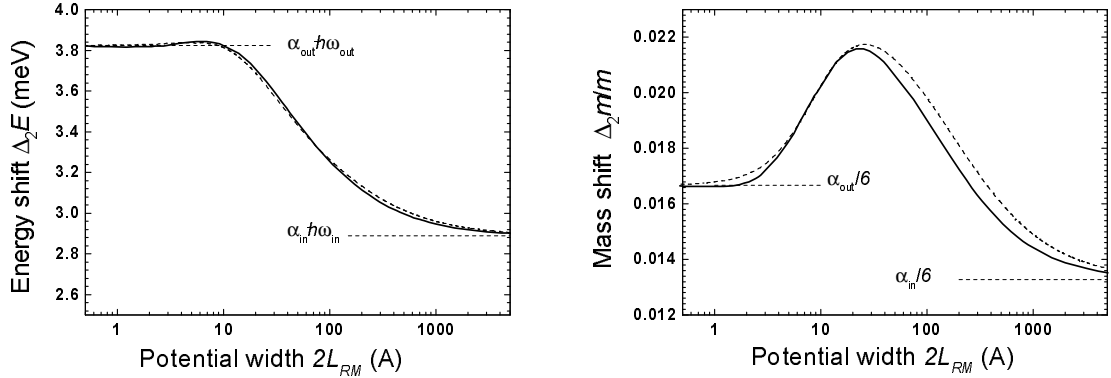}
  \vspace*{-19.0cm}
  \caption{Our results for the Rosen-Morse potential in comparison
with these for the rectangular potential (dashed line) of the same effective
width $L(L_{RM})$.
}
  \label{bog-fig2}\end{figure}

The polaron energy and effective mass are calculated for $x=0.3$. Peaks
are found for the effective mass at some potential widths, while
the energy demonstrates rather a smooth behavior between
the correct 3D-limits as is seen in Fig.~\ref{bog-fig1}.
As to the Rosen-Morse potential, the results are presented in
Fig.~\ref{bog-fig2} together with those for the rectangular potential
of the corresponding effective width. One can see an excellent coincidence
of the results obtained within the different techniques; clearly, this
fact increases their reliability. A comparison is also made with the results
of the papers\cite{Hai,Shi}, and the details are discussed in our
paper\cite{rosen}.

\section{Excitons}

Sampling the previous literature, most work has been done on
rectangular quantum wells with confinement potentials of type (\ref{eq3.01}).
The electron-hole
potential can be calculated as indicated above and was given e.g. in Ref.
\cite{Takagahara}.

To treat eigenvalue problems as the present one, we use tractable
decompositions of the Hamiltonian to generate lower bounds for the
ground-state energy. The basic idea is as follows: Assume we study
the Hamiltonian $H=p_z^2/2m + V_1(z) + V_2(z)$ to find its ground-state
energy $E$. Then we use the decomposition

\begin{eqnarray}
H_1=x {p_z^2 \over 2m} + V_1(z), \quad H_2=(1-x) {p_z^2\over 2m} +
V_2(z) ,
\qquad 0\leq x \leq 1  .
\label{eq_decomp}\end{eqnarray}

\noindent
If $E_1(x), E_2(x)$ are the corresponding ground-state energies
of $H_1, H_2$, then a lower bound for $E$ is: $E \geq \max_x (E_1(x) + E_2(x))$.

\begin{figure}[h]
  \hspace*{1.5cm}
  \epsfysize=3in \epsfbox{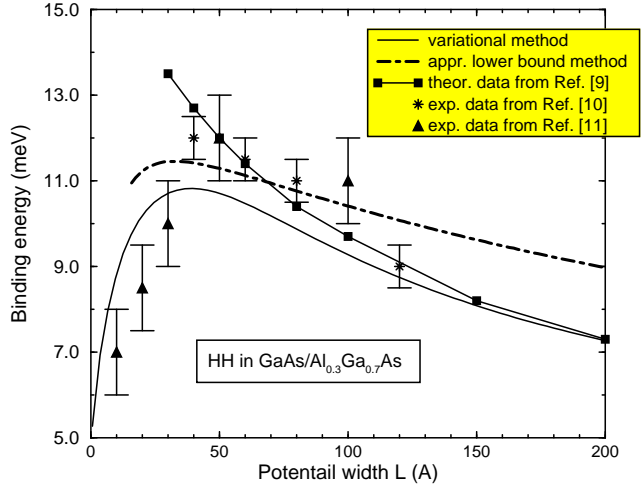}
  \caption{Comparison of results for the binding energy of an exciton in
          a rectangular quantum well as function of the well width.}
 \label{bog-fig3}\end{figure}

Upper bounds are produced by variational methods: The trial wave-function
used in our calculations had the form:

\begin{eqnarray}
\Psi(\vec r_{\perp}, z_1,z_2) =
\Phi_1(z_1)\Phi_2(z_2)e^{-a\sqrt{r_{\perp}^2+b (z_1-z_2)^2}}  ,
\label{eq-trail}\end{eqnarray}

\noindent
where $\Phi_i(z_i)$ are the ground-state eigenfunctions
of the free electron ($i=1$) or the hole ($i=2$) in the confining
potentials of the type (\ref{eq3.01}). Evidently, the variational
parameters $a,b$ can be used to fit 3D and 2D limiting cases.
If the masses can be assumed as constant over the heterostructure, these
methods can profitably be combined with functional-integral techniques.
Fig.~\ref{bog-fig3} shows our result \cite{Gerlach1} for
$Al_{0.3}Ga_{0.7}As/GaAs/Al_{0.3}Ga_{0.7}As$. in comparison with experimental
\cite{Gurioli,Voliotis} and previous theoretical results
\cite{Andreani}. Clearly, the maximum appears at a relevant width.

A second class of confinement potentials is of parabolic type, that is
\begin{equation}
V_i(z)= {m_iR_{\infty}^2 {\lambda}^2_i \over 2{\hbar}^2} z_i^2 ,
\end{equation}
where $\lambda_i$ denotes the dimensionless confinement strength, $R_{\infty}$
is the Rydberg unit, which was extracted for reasons of convenience.
To study the confinement-induced effects on the spectrum as accurately
as possible, we disregarded any parameter mismatch.
The quantity of interest is the diagonal element of the reduced density
operator, namely

\begin{equation}  \label{p1}
P_{\beta}({\bf{C}}):=<{\bf{C}}|tr_{Ph}e^{-{\beta}H}|{\bf{C}}>  .
\end{equation}

\noindent
In this formula ${\bf{C}}$ is an abbreviation for an arbitrary (but fixed)
set of the position coordinates of the particles involved. The
right-hand side of Eq. (\ref{p1}) can be represented as a functional
integral

\begin{equation}  \label{p2}
P_{\beta}({\bf{C}}) = Z_{Ph} \int \delta^{6}R\ e^{-S[{\bf{R}}]}  .
\end{equation}

\noindent
In Eq. (\ref{p2}) $Z_{Ph}$ is the free-phonon partition function, and
$S$ reads as follows:


\begin{eqnarray} \label{action}
S[{\bf{R}}] &:=&
  \int_0^{\beta} d\tau
  \left(\sum_{n=1}^2 \frac{m_n}{2} {\dot{\bf{R}}}_n^2(\tau) +
        U({\bf{R}}_1(\tau),{\bf{R}}_2(\tau))\right) \nonumber \\
&& -\sum_{n,n'=1}^2 \sum_{{\bf{k}}}
  \frac{g_{{\bf{k}},n} g_{{\bf{k}},n'}}{V}
  \int\limits_{0}^{\beta}\int\limits_{0}^{\beta}d\tau
  \,d\tau'\,G(\tau-\tau')\;
  e^{i{\bf{k}}[{\bf{R}}_n(\tau)-{\bf{R}}_{n'}(\tau')]}  .
\end{eqnarray}

\noindent
Within the functional integral (\ref{p2}), $\int \delta^{6} R....$ is
to indicate integration over all real, $6$-dimensional paths
${\bf{R}}(\tau)$, which start and end at the point ${\bf{C}}$.
The kernel function $G(\tau -{\tau}')$ is defined as

\begin{equation} \label{kernel}
G(\tau):={e^{\hbar\omega(\beta- |\tau|)} + e^{\hbar\omega|\tau|} \over
          2[e^{\beta \hbar \omega} - 1] } .
\end{equation}

It is well known that functional integrals of type (\ref{p2}) with an
action (\ref{action}) cannot be evaluated in analytical form. Starting
from the exact expression, we use variational procedures as in
Feynman's famous paper on polarons to find upper bounds on the
ground-state energy. The necessary input is a trial action, which is
accessible to a numerical treatment.

  \begin{figure}[h]
  \hspace*{1.5cm}
  \epsfysize=2.8in \epsfbox{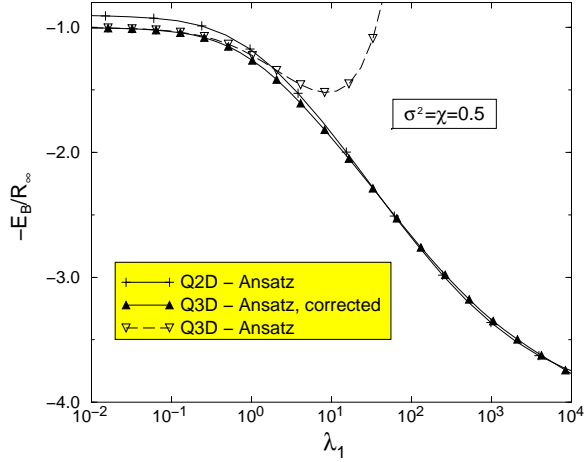}
  \caption{Binding energy of an exciton in a parabolic
quantum well as function of the electron confinement strength $\lambda_1$.
The comparison is made for different approaches described in
Ref.\protect\cite{Gerlach2}. The parameters $\sigma^2 = m_1/m_2$
and $\chi = \lambda_2/\lambda_1$ are fixed as indicated.}
 \label{bog-fig4}\end{figure}

\begin{figure}[t]
  \hspace*{1.5cm}
  \epsfysize=2.8in \epsfbox{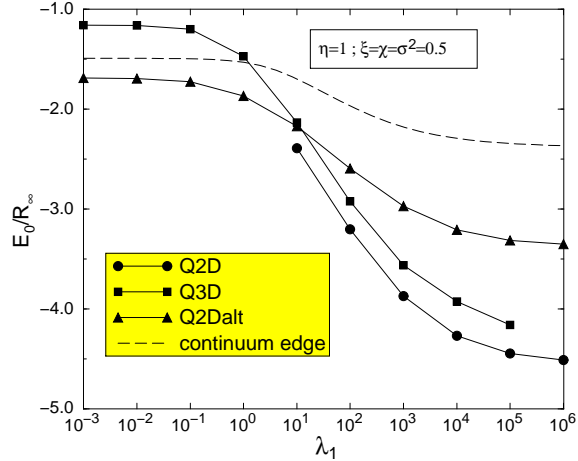}
  \caption{Ground-state energy of an exciton-phonon
system in a parabolic quantum well as function of the confinement electron
strength $\lambda_1$. The remaining parameters
$\eta = \protect\sqrt{R_{\infty}/\hbar\omega}$ and
$\xi = 1-\varepsilon_{\infty}/\varepsilon_0$
are fixed as indicated. In addition, an upper bound for the
energy of the continuum edge is shown.}
\label{bog-fig5}\end{figure}

The trial companions of the exact action (\ref{action}) were
combinations of oscillator trial actions for the center-of-mass and
the $z$-coordinate and three-dimensional (two-dimensional) Coulomb
potentials for the three-dimensional (two-dimensional in-plane) relative
coordinates. The corresponding results (see Ref. \cite{Gerlach2}) can be
found in the following figures and are denoted as quasi three-dimensional (Q3D) and
quasi two-dimensional (Q2D or Q2Dalt) ansatz. In Fig.~\ref{bog-fig4}
we neglect any phonon influence to demonstrate the smooth interpolation
of the limiting values $1 R_{\infty}$ and $4 R_{\infty}$ of the binding energy
(actually we plotted there the ground-state energy with the continuum
edge being subtracted, that is, the quantity $-E_B$).
Fig.~\ref{bog-fig5} shows results for the general case; we present data
for the ground-state energy as well as the continuum edge, which is the
reference for the binding energy and has to be calculated separately.

The results reported have been obtained in collaboration with M.\ Dzero
and J.\ W\"usthoff; we gratefully thank both of them.
We are indebted to J.~T. Devreese, V. Gladilin, H. Leschke, V.~M. Fomin,
F.~M.~Peeters, and E.~P. Pokatilov for useful discussions and remarks.
The support of Deutsche Forschungsgemeinschaft and the Germany-JINR
Heisenberg-Landau program is acknowledged.

\end{document}